\documentclass[graybox, envcountchap]{svmult}

\usepackage{mathptmx}        
\usepackage{helvet}          
\usepackage{courier}         

\usepackage{makeidx}         
\usepackage{graphicx}        
\usepackage{multicol}        
\usepackage[bottom]{footmisc}

\makeindex             

\usepackage{amsmath}
\usepackage{amssymb}
\usepackage{lscape}
\usepackage{multirow}

\def\gapp{\ifmmode\stackrel{>}{_{\sim}}\else$\stackrel{<}{_{\sim}}$\fi}
\def\gsim{\lower.5ex\hbox{\gtsima}}
\def\gtsima{$\; \buildrel > \over \sim \;$}
\def\lapp{\ifmmode\stackrel{<}{_{\sim}}\else$\stackrel{<}{_{\sim}}$\fi}
\def\lsim{\lower.5ex\hbox{\ltsima}}
\def\ltsima{$\; \buildrel < \over \sim \;$}

\newcommand\apgt{\ {\raise-.5ex\hbox{$\buildrel>\over\sim$}}\ }
\newcommand\aplt{\ {\raise-.5ex\hbox{$\buildrel<\over\sim$}}\ }
\begin{document}
\pagestyle{empty}
\frontmatter

\include{dedic}
\include{foreword}
\include{preface}

\mainmatter

\setcounter{chapter}{6}
\title{Mass Transfer by Stellar Wind}
\author{Henri M.J. Boffin}
\institute{Henri M.J. Boffin \at ESO, Alonso de Cord\'oba, 3107, Vitacura, Santiago 190001, Chile,\\ \email{hboffin@eso.org}
}
%
%
\maketitle
\label{Chapter:Boffin}

\abstract*{I review the process of mass transfer in a binary system through a stellar wind, with an emphasis on 
systems containing a red giant. I show how wind accretion in a binary system is different
from the usually assumed Bondi-Hoyle approximation, first as far as the flow's structure is concerned, but most importantly, also for the mass accretion and specific angular momentum loss. This has important implications on the evolution of the orbital parameters. I also discuss the impact of wind accretion, on the chemical pollution and change in spin of the accreting star. The last section deals with observations and covers systems that most likely went through wind mass transfer: barium and related stars, symbiotic stars and central stars of planetary nebulae (CSPN). The most recent observations of cool CSPN progenitors of barium stars, as well as of carbon-rich post-common envelope systems, are providing unique constraints on the mass transfer processes.}

\section{Stars in couple}
\label{Boffinsec:1}

A large fraction of stars are found in binary or multiple systems\index{binary star}. This fraction is dependent on the mass of the primary (i.e., the more luminous of the two stars), and is about 40--50\% for solar-like stars \cite{bofRaghavan2010} and 70\% for A stars \cite{2014MNRAS.437.1216D}, while it is even higher for more massive stars (which are not considered in this chapter). When close enough, the stars in a binary system may interact through tidal forces, mass and angular momentum transfer. What is ``close enough''? To answer this requires to introduce  the concept of
Roche lobe\index{Roche lobe}.

To do this, we consider the simplified case of a system containing two stars, of mass $M_1$ and $M_2$, \emph{in a circular orbit} and with a separation $a$. Kepler's third law tells us that the orbital period $P$ is then given by
\begin{equation}
P = \frac{2\pi}{\omega} = 2\pi \sqrt{ \frac{ a^3}{G (M_1+M_2)}  },
\end{equation}
where $\omega$ is the orbital angular velocity and $G$ is the gravitational constant. 
If we assume that the two stars are points and that \emph{ their rotation is synchronised with the orbital motion}, the potential $\Phi$ in the rotating system with the centre of mass at the origin is given by
\begin{equation}
\label{Eq:Roche}
\Phi = -\frac{q}{1+q}\frac{1}{r_1}  -\frac{1}{1+q}\frac{1}{r_2} -\frac{1}{2}~(x^2+y^2).
\end{equation}
Here, $x$ is along the line joining the two stars, while $y$ is perpendicular to $x$ in the orbital plane and $z$ is perpendicular to the orbital plane, $q$ is the mass ratio, $q=M_1/M_2$, and $r_i$ is the distance to star $i$, given by 
 \begin{equation}
r_i = \sqrt{ (x - x_i)^2 + y^2 + z^2  },
\end{equation}
and $x_1 = 1/(1+q)$ while $x_2 = -q/(1+q)$. In the above, we have assumed the usual normalisation, $a=G=M_1+M_2=1$.

\begin{figure}
\begin{center} 
\includegraphics[width=119mm]{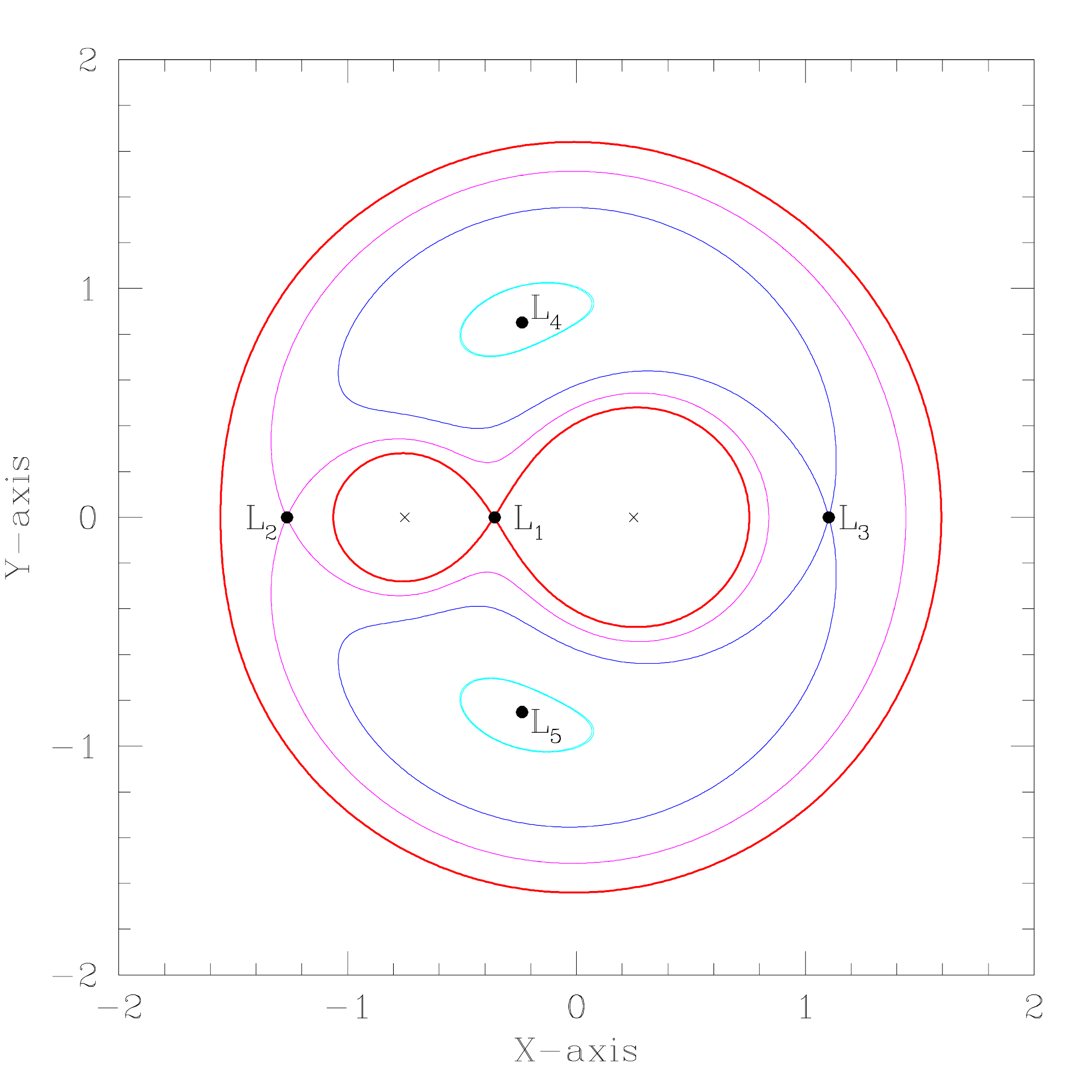}
\caption{Equipotentials in the orbital plane of a binary system with $q=3$. The rotating reference's origin is the centre of mass of the system, and the positions of the two stars are indicated by crosses. The Lagrangian equilibrium points, $L_1$ to $L_5$, are also indicated.}\label{fig:BoffinRoche} 
\end{center}
\end{figure}

Equipotentials \index{Equipotentials} as represented by Eq.~\ref{Eq:Roche} correspond to loci of constant pressure, which means that a star in a binary system will take their shape. An example of the resulting equipotentials in the orbital plane ($z=0$) for a binary with a mass ratio of three is shown in Fig. \ref{fig:BoffinRoche}. There are five equilibrium points where the gradient of the potential is zero, and those points are called the Lagrangian points ($L_1$, $L_2$ and $L_3$ are aligned on the $x$-axis and are unstable). 
The equipotential passing through the inner Lagrangian point $L_1$ is essential in the study of binary stars. It is called the Roche equipotential and it defines, around each of the two stars, a pear-shaped region, called  \index{Roche lobe}\emph{Roche lobe}. This region represents the maximum radius a star can reach before there will be  mass transfer\index{mass transfer}. 
If both stars are well within their Roche lobe, the system is called \emph{detached}. In this case, both stars will mostly keep a spherical shape and there is a limited interaction between them. If either the separation between the two stars decreases (and thus the Roche lobes as well) or if the radius of the stars increases (due to stellar evolution), one of the star -- generally the most massive as this is the one that will evolve faster 
-- will fill its Roche lobe, taking a pear shape. The system is then \emph{semi-detached}, and mass transfer takes place: material will flow from the Roche lobe filling star to its companion via the inner Lagrangian point $L_1$. Such mass transfer is called \emph{Roche lobe overflow} (RLOF)\index{Roche lobe overflow}. In some particular cases, e.g. in W UMa\index{W UMa star}, the mass transfer results in both stars  filling their Roche lobes and the system is then called \emph{in contact}, both stars being surrounded by a \index{common envelope}\emph{common envelope} \cite{bofko55}. Note that when the star fills a large fraction of its Roche lobe, tidal effects become very strong, and the star's rotation become synchronised with the orbital motion on a short timescale, while the orbit will quickly circularise as well. This justifies the assumptions made in the definition of the Roche lobe, although it is important to realise that they may not be always fulfilled. 

The position of the $L_1$ point is given by \cite{PlavecKratochvil64}:
\begin{equation}
x_{L_1} = \frac{1}{1+q} - (0.5 + 0.227~\log{q}).
\end{equation}
One also usually defines the Roche lobe radius, $R_L$ as the radius of a sphere whose volume equals the volume of the Roche lobe itself. Some useful approximations for the Roche lobe radius around star 1 were derived by \cite{pacz71} from the values tabulated by \cite{ko59}
\begin{eqnarray}
R_L =& 0.462~a \left( \frac{q}{1+q} \right) ^{\frac{1}{3}}   &~~~~~~~~~~q < 0.523 \\
R_L =& a~(0.38 + 0.2 \log q)  &~~~~~0.523 < q < 20,
\end{eqnarray}
and, later, by \cite{Eggleton83}
\begin{equation}
R_L = \frac{0.49~q^{\frac{2}{3}}}{0.6~q^{\frac{2}{3}} + \ln (1+q^{\frac{1}{3}})} ~a.
\end{equation}
The Roche lobe radius around star 2 is obtained by replacing $q$ by 1/$q$.

Here, we will only deal with systems containing low- and intermediate-mass stars, i.e., stars in the range $0.8 < M < 8$~M$_\odot$, as these are the only one relevant to blue straggler stars. The typical maximum radius such a 
star will reach on the first giant branch (RGB)\index{red giant} is about 200 R$_\odot$ for stars below about 1.3--1.5 M$_\odot$, but only 30~R$_\odot$ to $\sim100$~R$_\odot$ for stars between 2 and 3~M$_\odot$. On the other hand, the radius such stars can reach on the asymptotic giant branch (AGB)\index{asymptotic giant branch} is of several hundreds of solar radii, especially during the last phases, when the AGB undergoes thermal pulses. In these cases, the maximum radius can reach up to 800~R$_\odot$ for stars above 3~M$_\odot$. Thus, except for the less massive stars, a star can avoid to fill its Roche lobe if the initial orbital period is above, say, 50 to 200 days\index{orbital period}. For the less massive stars, Roche lobe overflow could in principle happen for periods up to 1000 days! And in all cases, RLOF could happen on the AGB -- at least in the very last phases of this evolution -- for orbital periods up to 10--20 years!
We will see later, however, that even in this case, it is most likely that prior to the RLOF, there will be an extensive phase of wind mass transfer.

\begin{figure}
\begin{center} 
\includegraphics[width=119mm]{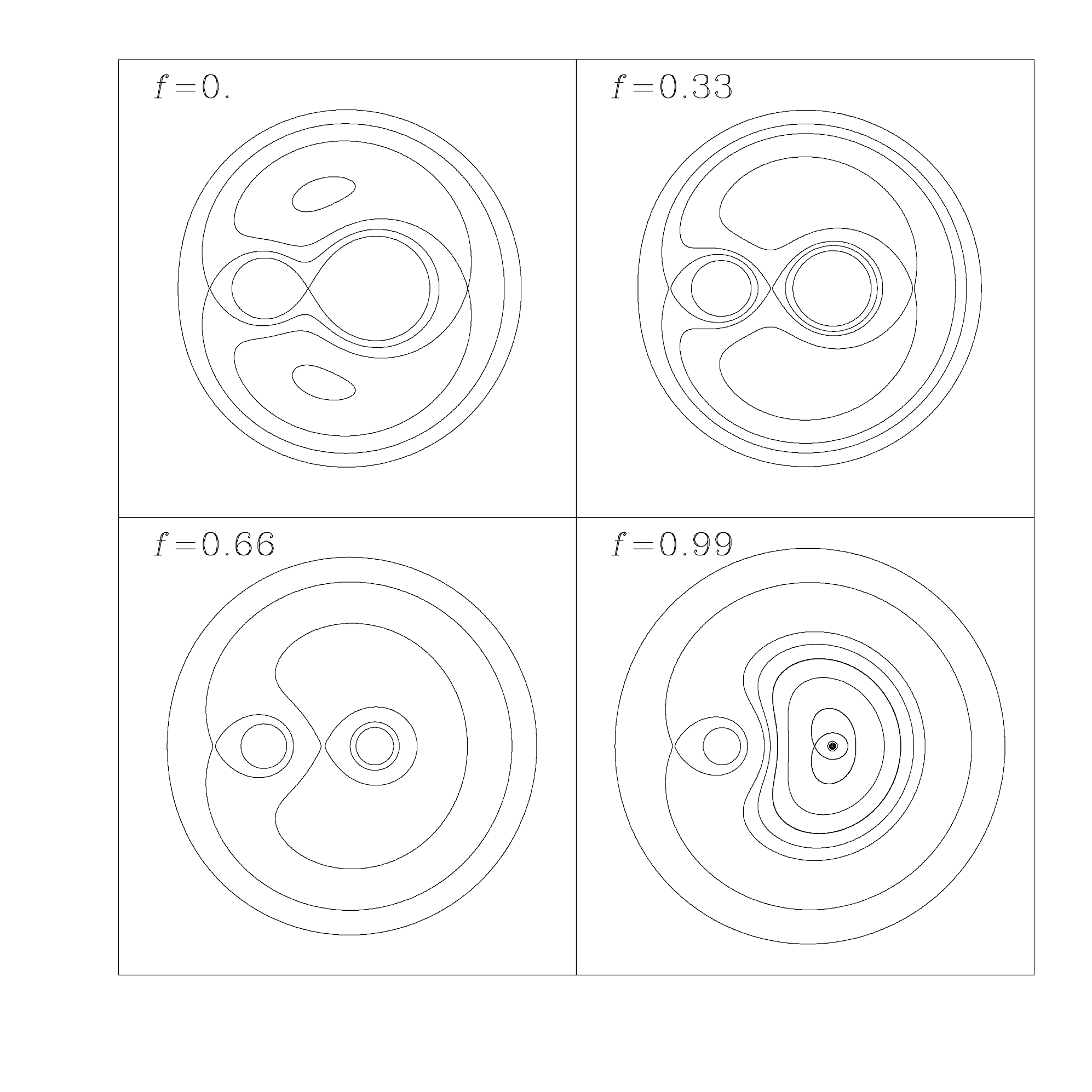}
\caption{Same as Fig.~\ref{fig:BoffinRoche} for the case of a reduced effective gravity due to the mass loss by stellar wind. The gravity of the primary is multiplied by a factor $1.-f$, where $f$ takes the values 0, 0.33, 0.66, and 0.99. As can be seen, for large values of $f$, the shapes of the Roche lobes change dramatically, and one can no more define the usual Lagrangian points. This will affect the way matter is transferred from one star to the other. }\label{fig:BoffinRocheRed} 
\end{center}
\end{figure}

The description above may need some modification when dealing with red giants. Such stars will indeed lose mass (see below) through stellar wind. The exact way this happens in AGB stars is still far from understood, although it generally assumes that pulsation is pushing material far enough away from the star for dust to condense and radiation pressure onto these dust grains will lead to the mass loss. This implies that the effective gravity of the star has been reduced -- multiplied by $(1-f)$, where $f$ is an unknown factor between 0 and 1. This reduced gravity will change the shape of Roche lobes as already shown by \cite{Schu72} and more recently by \cite{Der09}. Figure~\ref{fig:BoffinRocheRed} shows the effect this has on the equipotentials and thus on the way mass will be transferred. Whether this is actually taking place in binary systems is still open to debate.

\section{Wind mass transfer}
\label{Boffinsec:2}

In this chapter, we are concerned with the case of detached binaries, that is, both stars are well within their Roche lobe. 
The case of Roche lobe overflow is covered in Chap. 8.

Being detached does not imply that there is no mass transfer. Indeed, stars undergo regular mass loss, via a stellar wind\index{stellar wind}, and in such case, a fraction of the mass lost may also be captured by the companion, which will thereby gain mass and angular momentum. The Sun is losing mass at a rate of $10^{-14}$~M$_\odot$yr$^{-1}$ -- a value too small to be of any interest in the evolution of binary stars. At such a rate, over 10 billion years only $10^{-4}$~M$_\odot$ would have been lost and if the Sun had a binary companion, only a fraction of it would have been accreted, hardly worth mentioning. But there are phases in the lives of solar-like stars where the mass loss by wind could be much stronger. Similarly, high mass, i.e. O and B, stars, experience much higher mass loss rates by wind\index{mass loss}.

Red giants with luminosity of 50--200~L$_\odot$ have mass loss rates of a few $10^{-10}--10^{-8}$~M$_\odot$yr$^{-1}$, while AGB stars will lose more, and in their final stages, as carbon stars, they may loose between $10^{-7}$~M$_\odot$yr$^{-1}$ and $10^{-4}$~M$_\odot$yr$^{-1}$. Such massive mass loss will have quite an impact on the evolution of the binary system.

\subsection{The Bondi-Hoyle-Lyttleton model}\index{wind accretion}\index{Bondi-Hoyle accretion}
\begin{figure}
   \begin{center}
   \includegraphics[width=119mm]{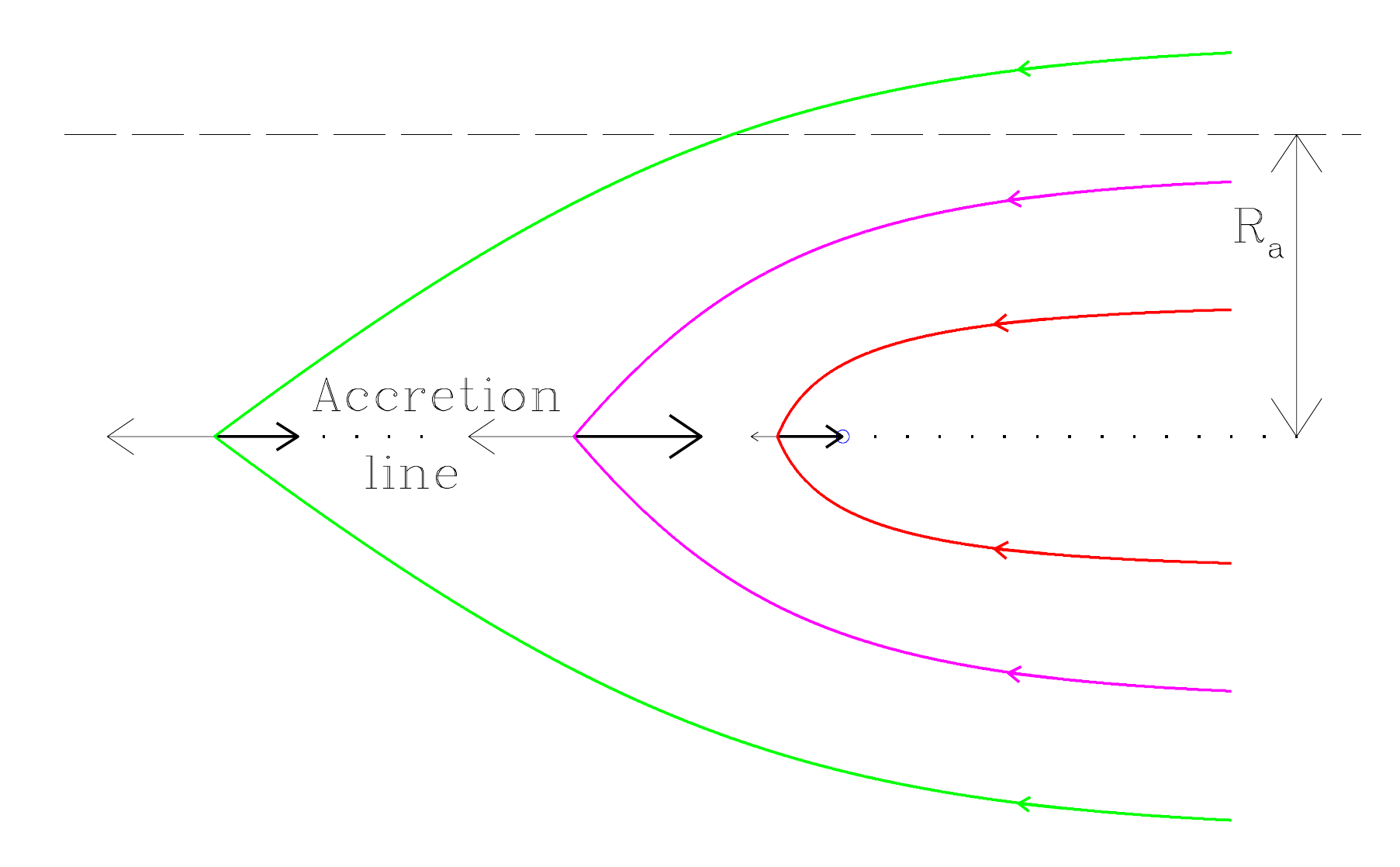} 
   \caption{Hoyle-Lyttleton representation of the wind accretion process: a star (indicated by the blue circle) moves inside a cloud of cold gas. The accretion radius is defined as the largest distance from the star, where the $x$-velocity behind the star (and the sole velocity after collision; indicated by the arrows) is equal to the escape velocity (represented by the bold arrows). All material within the accretion radius is supposed to be accreted by the star.}
   \label{Boffig:hl}
   \end{center}
\end{figure}

The simplest problem to consider in the framework of wind accretion is that of a star of mass $M$ moving with velocity $v_\infty$ in a homogeneous, uniform and inviscid cloud of density $\rho_\infty$. Far away from the object, the pressure forces will be small and the matter will follow a free Keplerian orbit. Approaching the gravitational mass, the particle trajectories are deflected and, behind the object, a region forms where particles collide and increase the density. In this region, pressure effects become important. This problem was first considered by \cite{HL39} who studied the effect of interstellar matter accretion by the Sun on the Earth climate! They assumed a cold gas, and therefore neglected any pressure effect. In
this model, gravitationally deflected material passing on one
side of the star collides with material passing on the opposite side,
thereby cancelling its transverse velocity and forming an ``accretion line'' behind the star. The material from this accretion
line that has a velocity below the local escape velocity from the star
will be accreted (Fig.~\ref{Boffig:hl}), leading to a mass accretion rate\index{accretion rate} $\dot M_{HL}$:
\begin{equation}
\dot M_{HL} = \pi R_a^2 \rho_\infty v_\infty,
\label{eq:HL}
\end{equation}
where the accretion radius $R_A$ is given by
\begin{equation}
R_a = 2~\frac{GM}{v^2_\infty}.
\label{eq:RA}
\end{equation}
This result is only valid for hypersonic flows, i.e., $v_\infty \gg c_\infty$, the sound velocity at infinity.
Later, Bondi \& Hoyle \cite{BH44} included some limited pressure effects (making the accretion line, an ``accretion column''), and showed that Eq.~\ref{eq:RA} is an upper limit, and that the actual accretion rate is given by 
$$\dot M = \alpha \dot M_{HL},$$
where $\alpha$ is an efficiency parameter of the order of unity. Physically, $\alpha$ represents the location of the stagnation point in units of the accretion radius. In the Bondi \& Hoyle formalism, this position cannot be determined exactly.

\begin{figure}
   \begin{center}   \includegraphics{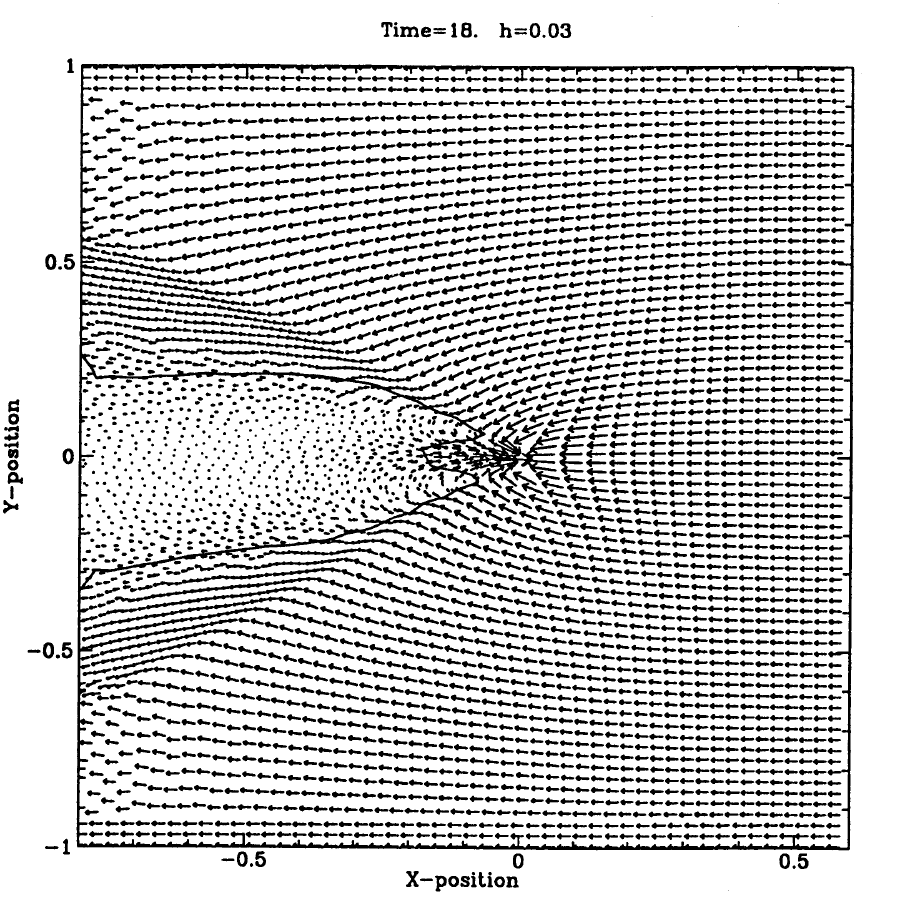} 
   \caption{Two-dimensional numerical simulation of a Bondi-Hoyle accretion at Mach 3. The arrows illustrate the velocity field, while the thick line represents the contour of Mach=1. A shock has been formed but there is still a stagnation point clearly visible, defining the accretion radius. Reproduced with permission from Fig. 1 of Boffin \& Anzer \cite{BA94}.}
   \label{Boffig:BA}
   \end{center}
\end{figure}

Bondi \cite{Bondi52} studied the complementary case of a pressure dominated flow, namely a stationary, spherically symmetric accretion with $v_\infty=0$. The accretion rate in a steady state is not uniquely
determined but the model does provide a maximum accretion rate $\dot
M_{B}$:
\begin{equation}
\dot M_{B} = \lambda \pi R_B^2 \rho_\infty c_\infty,
\label{eq:B}
\end{equation}
where the Bondi radius is given by Eq.~\ref{eq:RA} but with the
sound speed at infinity $c_\infty$ replacing $v_\infty$. Here $\lambda$
ranges between 0.25 and 1.12, with the exact value 
depending on the
polytropic index\index{polytropic index} of the gas. Bondi then proposed an interpolation formula
for the intermediate cases, which is the most widely used  Bondi-Hoyle accretion rate:
\begin{equation}
\dot M_{BH} = \alpha \pi R_{A}^2 \rho_\infty v_\infty 
              \left(\frac{{\cal M}_\infty^2} 
                    {1 +  {\cal M}_\infty^2}\right)^{3/2},
\label{Eq:BH}
\end{equation}
where ${\cal M}_\infty = v_\infty/c_\infty$ is the Mach number.

The theoretically predicted accretion rate $\dot M_{BH}$ in the
Bondi-Hoyle model has been extensively tested against numerical
simulations by, e.g.,  \cite{BA94,Hunt71,2006PThPh.116.1069I,Matsuda87,Matsuda92,Ruffert94,Shima98}. These show that 
the accretion line is only present transiently   as, soon, pressure effects will create a shock upstream of the star (Fig.~\ref{Boffig:BA}). Still, the theoretical and numerical rates
agree to within 10--20\%.

When dealing with binary systems, Boffin \& Jorissen replaced $v_\infty$ by $\sqrt{v_w^2+v_{\rm orb}^2}$, where $v_w$ is the wind speed and $v_{\rm orb}$ is the orbital velocity. The density $\rho_\infty$ is  computed from assuming a spherical mass loss from the primary and averaging over one orbit, $\dot M_w=4 \pi \rho_\infty a^2 \sqrt{1-e^2} v_w$. 
This then leads to a mass transfer rate, $\beta$:
\begin{equation}
\beta = -\frac{\dot M_{\rm acc}}{\dot M_w} =  \alpha \frac{1}{\sqrt{1-e^2}}\left(\frac{q}{1+q}\right)^2 \frac{v_{\rm orb}^4}{v_w~(v_w^2+v_{\rm orb}^2+c^2)^{3/2}}.
\label{Eq:BHbin}
\end{equation}

One can see that in the case of a fast wind (the only case when this equation is formally valid), $v_w^2 \gg v_{\rm orb}^2+c^2$, we have 
\begin{equation}
\beta \propto  \left( \frac{v_{\rm orb}}{v_w} \right)^4,
\end{equation}
while in the opposite case,  
\begin{equation}
\beta \propto    \frac{v_{\rm orb}}{v_w},
\end{equation}
which could lead to the unphysical case $\beta > 1.$

\begin{figure}
\begin{center} 
\includegraphics[width=99mm]{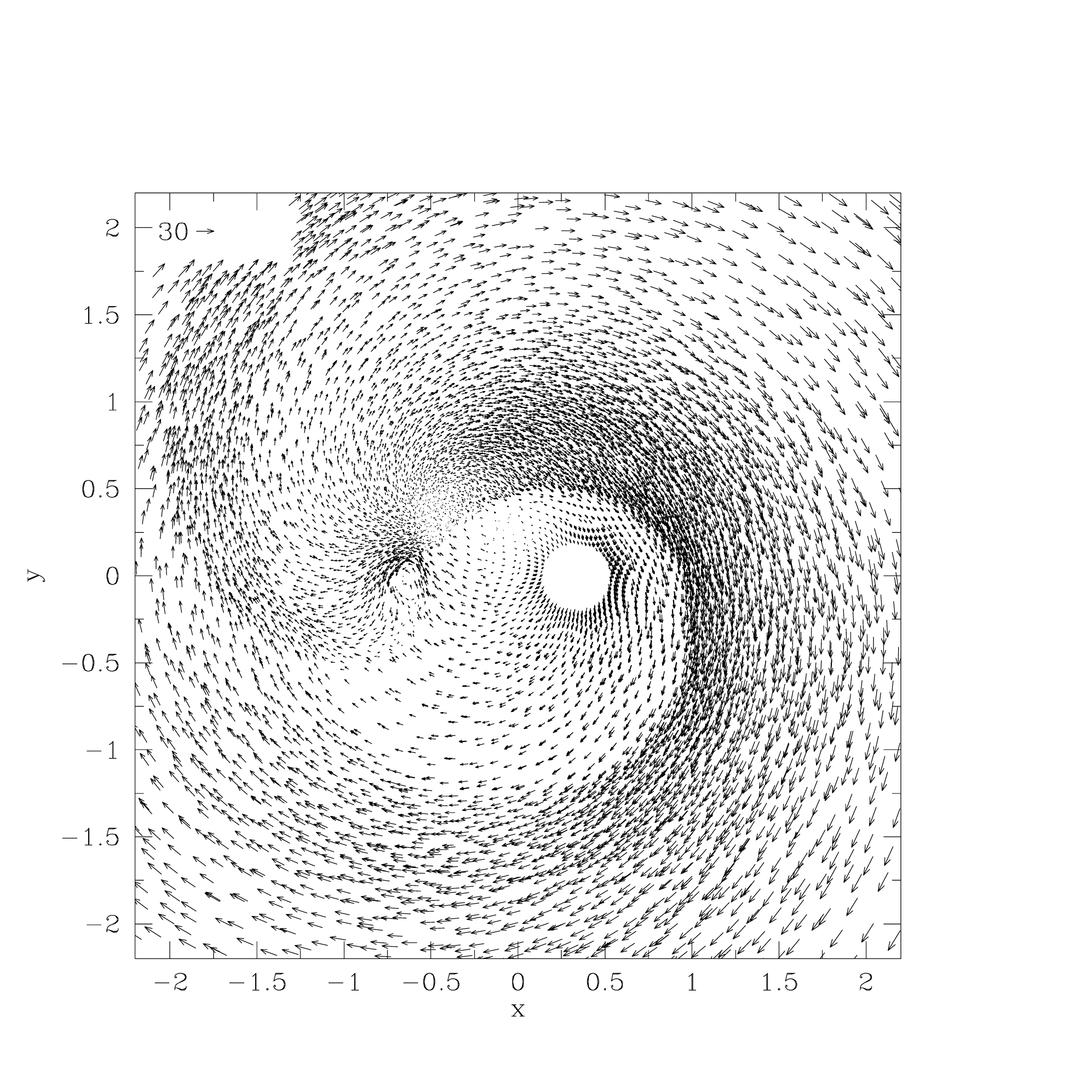}
\caption{Numerical simulations of mass transfer in a 3 AU binary system, where the primary looses mass through a slow wind. A clear spiral structure is seen in the orbital plane, and the flow is therefore much different from what one expects for a normal Bondi-Hoyle flow, due to the effect of the centrifugal force. Reproduced from Theuns et al. \cite{TBJ96}.}\label{fig:BoffinTBJ} 
\end{center}
\end{figure}


\section{Wind accretion in binary systems}

The Bondi-Hoyle formulation (Eq.~\ref{Eq:BHbin}) is very widely used -- in particular in population synthesis calculations, despite the fact that, 
while it may be valid when the orbital velocity is much smaller than the wind velocity, i.e. in $\zeta$ Aurigae and O-type binaries, it has been shown since two decades that it does not apply to binary systems containing a red giant. 
In such systems indeed, and unless the system is very wide, the wind velocity (5--30 km/s, depending on the evolutionary stage) of a red giant is smaller than or of the order of magnitude of the orbital velocity, and the binary motion becomes important in determining the flow structure. This, has in turn, also a big impact on the mass accretion rate.

This was first shown by Theuns, Boffin \& Jorissen \cite{BTJ94,TBJ96,TJ93} who clearly demonstrated that the flow structure of a mass-losing AGB star in a 3 AU system was very different from what one expects for a Bondi-Hoyle accretion flow (Fig.~\ref{fig:BoffinTBJ}), with the presence of a large spiral. Such spirals have since then been detected observationally by HST in the carbon star\index{carbon star} AFL3068\index{AFL3068} \cite{Sahaietal2006} and with ALMA in R Scl\index{R Scl} \cite{Maercker12}.  
Such a different flow structure also impact the mass accretion rate and Theuns et al. \cite{TBJ96} found a mass accretion rate $\beta$ of the order 1\% in the
adiabatic case, and 8\% for the isothermal case. In the adiabatic case, \emph{this is about ten to twenty times smaller than the theoretical estimates} based on the
Bondi-Hoyle prescription, i.e. $\alpha \sim 0.05-0.1$.

\begin{figure}[htbp]
\begin{center} 
   \includegraphics[width=119mm]{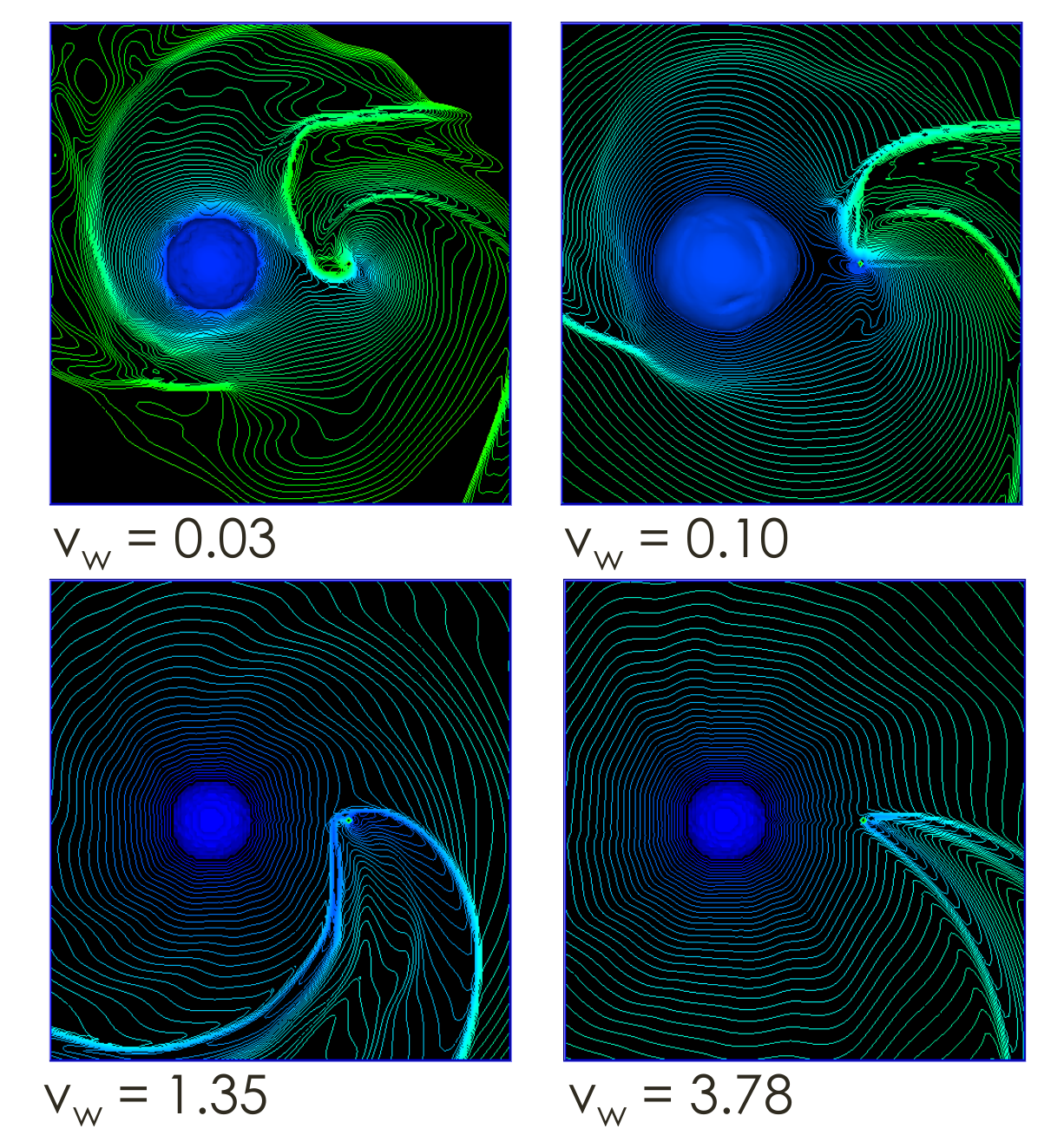}
   \caption{Numerical simulations of mass transfer in a binary system, for various wind speed, $v_w$, in units of the orbital velocity. The low velocity case shows resemblance to a Roche lobe overflow, although mass also escape through the outer Lagrangian point, while the higher velocity case are reminiscent of $\zeta$ Aurigae systems. Adapted from Nagae et al. \cite{Nagae04}.}
\label{nagae}
 \end{center}
  \end{figure}

\begin{figure}[htbp]
\sidecaption
   \includegraphics[width=7.5cm]{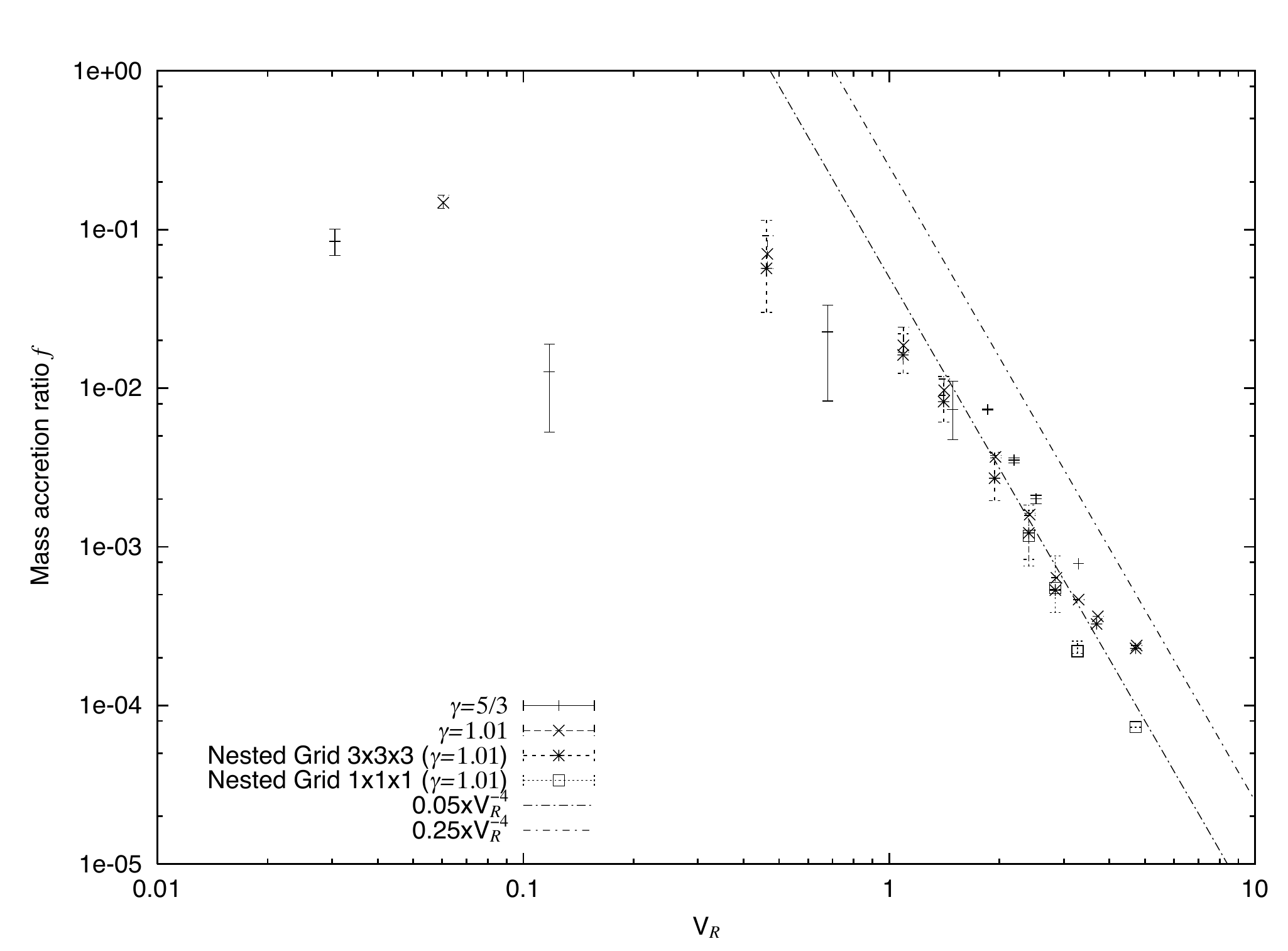}
   \caption{Log-log plot of the mass accretion ratio as a function of the 
            wind velocity at the position of the Roche lobe. The 
            upper line shows a simple Bondi-Hoyle-Lyttleton formula with $\alpha=1.$ 
	    The lower dashed line shows another 
            empirical formula $f = 0.05 v_R^{-4}$, i.e. 20\% the Bondi-Hoyle value. Reproduced with permission from Nagae et al. \cite{Nagae04}.}
\label{log-log}
   \end{figure}

A more extensive study was done by \cite{Nagae04} who explored the full parameter space, varying the wind velocity with respect to the orbital speed, in the case of $q=1$ (Fig.~\ref{nagae}). They characterised the flow by the mean normal velocity of the wind on the critical Roche surface of the mass-losing star, $v_R$. When $v_R < 0.4 v_{\rm orb}$, they obtained Roche-lobe over-flow (RLOF), while for $v_R > 0.7 v_{\rm orb}$ they observed wind accretion. Very complex flow patterns in between these two extreme cases were found. They also found that mass accreted by the accreting star is roughly \emph{a factor 5 smaller than what would be naively deduced from the Bondi-Hoyle-Lyttleton formalism} (Fig.~\ref{log-log}).

For low $v_R$, they constructed an 
    empirical formula for the mass accretion ratio given by 
    \begin{equation}
    \beta = 0.18\times 10^{-0.75 v_R/v_{\rm orb}}.
    \end{equation}

  For larger $v_R$ they obtain another empirical formula for the mass
   accretion ratio  
       \begin{equation}
       \beta = 0.05~\left( \frac{v_{\rm orb}}{v_R} \right)^{4},
  \end{equation}
indicating that in the case $q=1$ they studied, $\alpha=0.2$ in Eq.~\ref{Eq:BHbin}.  

\subsection{Chemical pollution}
The accretion of material from an evolved star, whose surface composition has been modified from its initial composition (which we can safely assume is also the one of its companion), will lead to  chemical pollution. This is for example the case if mass is transferred from an AGB star which already went through thermal pulses and is enriched in carbon and s-process elements. As shown by \cite{BJ88,TBJ96}, the pollution of the envelope of the accreting star depends on 
\begin{itemize}
\item the
amount of mass $\Delta M_2$ accreted from the wind, 
\item the dilution
factor, ${\cal F}$, of the accreted matter in the envelope of mass $M_{2,env}$, i.e., 
${\cal F}=\Delta M_2/(M_{2,env}+\Delta M_2)$, and 
\item the difference in chemical composition between the accreted matter and
that of the envelope.  
\end{itemize}
Note that one can generally safely assume that the accreted matter has
been fully mixed in the envelope.
The overabundance $f_i$ of element $i$ in the envelope of the
accreting star (i.e., the ratio between the abundance after completion of
the accretion and mixing processes, and the abundance in the
primordial envelope) is related to its overabundance $g_i$ in the wind
(i.e., in the AGB atmosphere) through the relation
\begin{equation}
\label{BoffinEq:overab}
f_i = \frac{         g_i  \Delta M_2 + M_{2,env} }{ \Delta M_2 + M_{2,env} } 
   \equiv g_i {\cal F} + (1 -  {\cal F}).
\end{equation}

\subsection{Orbital parameters evolution}

The mass lost from the system, carrying with it angular momentum, and the mass
transferred from one star to
the other, will lead to a change in the orbital parameters.
Boffin \& Jorissen \cite{BJ88} presented a first analysis of this changes, based on the formalism of \cite{Huang56}. 
This was revisited by \cite{TBJ96} who derived:
\begin{eqnarray}
\frac{\dot a}{a} & = & -\phantom{2}{\dot M_1+\dot M_2\over M_1+M_2}
   - 2 {F_y\over M_2 v_{\rm orb}},
\label{boffineq:dAdt}\\
{\dot P\over P} & = & -2 {\dot M_1+\dot M_2\over M_1+M_2} 
     -3 {F_y\over M_2 v_{\rm orb}},
\label{boffineq:dPdt}\\
{\dot e\over e} & = &
-{1\over 2} {\dot M_2\over M_2} +{3\over 2} {F_y\over M_2 v_{\rm orb}},
\label{boffineq:dedt}
\end{eqnarray}
where $F_y$ is the tangential component of the total force, 
including gravity and accretion of linear momentum. 

Karakas, Tout \& Lattanzio \cite{KTL00} propose a different set of equations
\begin{equation}
\frac{\dot a}{a} = -\frac{\dot M_1}{M_1+M_2}-\left(\frac{1+e^2}{M_1+M_2}+\frac{2-e^2}{M_2}\right)
\frac{\dot M_2}{1-e^2},
\label{boffineq:karda}
\end{equation}

\begin{equation}
{\dot e\over e} = -\dot M_2 \left(\frac{1}{M_1+M_2}+\frac{1}{2 M_2}\right).
\label{boffineq:karde}
\end{equation}

In the approximation that little mass is gained, i.e. $\dot M_2\approx0$, the above equations lead to a
widening of the orbit and no change in eccentricity. This is what lead Boffin et al. \cite{BCP93} to propose that barium stars have evolved from normal red giants at constant eccentricity, and only their orbital period increased.
The resulting change in orbital period according to Eq.~\ref{boffineq:karda} for no mass accretion is thus
\begin{equation}
\frac{P}{P_0} = \left(  \frac{M_{1,0}+M_2}{M_1+M_2}\right)^2,
\label{boffineq:varpe}
\end{equation}
where it is assumed that $M_{2,0}=M_2$. 
Thus, assuming an initial binary system containing a 3~M$_\odot$ AGB star and a 1.5~M$_\odot$ companion with an initial period of 3 years (to  avoid Roche lobe overflow by the AGB), leads to a final orbital period of 13.77 years when the AGB became a 0.6~M$_\odot$ white dwarf. It is therefore difficult to explain the orbital periods below 10 years by such mechanisms. 
To explain shorter systems, one would thus need either to have consequent mass accretion or to have extra angular momentum loss (see Sect.~\ref{bofsec:Jaha}). 
If there is mass accretion, then the variation of the period is given by (assuming $e=0$)
\begin{equation}
\frac{P}{P_0} = \left(  \frac{M_{1,0}+M_{2,0}}{M_1+M_2}\right)^{2} \left(  \frac{M_{2,0}}{M_2}\right)^2,
\label{boffineq:varpeacc}
\end{equation}
so that starting from initial period of 3 years, we have a final period of 5.2, 3.5, 2.4 and 1.7 years, when accreting 20\%, 30\%, 40\% or 50\%, respectively.
This may be particularly relevant for blue straggler stars that have most likely accreted a large fraction of their mass (see Chap. 3). Those seen in the open cluster NGC 188 have current orbital period around 1000 days. They could have thus been the result of wind mass transfer coming from wider systems, if they accreted enough mass.

\subsection{Spin-up of accretor}
\label{bofsec:rot}
The companion star may accrete spin angular momentum from the wind,
which may alter its rotational velocity\index{rotational velocity}\index{stellar rotation}. Packett \cite{Packett81} showed that a
star needs to accrete only a few percent of its own mass from a
keplerian accretion disc\index{accretion disc} to be spun up to the equatorial centrifugal
limit, essentially because stellar moments of inertia are generally
much smaller than $MR^2$. 
Theuns et al. \cite{TBJ96} found that in their simulations of wind accretion in a binary system there is a
net accretion of spin momentum, in such a way
that the accreting star tends to be spun up if it is in synchronous
rotation to begin with (see also \cite{1996MNRAS.279..180J}). 
This could explain the rapid rotation\index{stellar rotation} of some barium stars, such  as HD~165141 \cite{Jo95}, as well as the barium-enriched central stars of planetary nebulae (see Sect.~\ref{bofsec:ring}).

\subsection{Angular momentum loss}\index{angular momentum}
\label{bofsec:Jaha}

As shown above, an isotropic mass loss from the mass-losing star without any accretion will  lead to an increase in the semi-major axis. When the angular momentum loss is large enough,
however, Hachisu et al. \cite{Hachisu99} showed that the system can shrink. They use this property to provide a wider channel for symbiotic\index{symbiotic star} systems to become Type Ia supernovae\index{Type Ia supernova}. The symbiotic binary system they consider consists of a white dwarf and a mass-losing red giant. If the white dwarf accretes enough of the wind from the giant, it could reach the Chandrasekhar limit and explode as a Type Ia supernova. An essential part of the validity of this single-degenerate channel resides in the fact that the required close white dwarf -- red giant binary can form from a very wide binary. They reach their conclusion after noticing that a system, where one of the components has a wind whose velocity is smaller or on the order of the orbital velocity, takes away the orbital angular momentum effectively, hence shrinking the orbit.
Jahanara et al. \cite{Jaha05}  made detailed three-dimensional hydrodynamic calculations of mass transfer in an interacting binary system in which one component undergoes mass loss through a wind, for various values of the mass ratio. They were particularly interested in looking at the amount of specific angular momentum of gas escaping the system ($l_w$) in order to see if the scenario of  Hachisu et al. was valid.    

      \begin{figure}[htbp]
\begin{center} 
   \includegraphics[width=11.5cm]{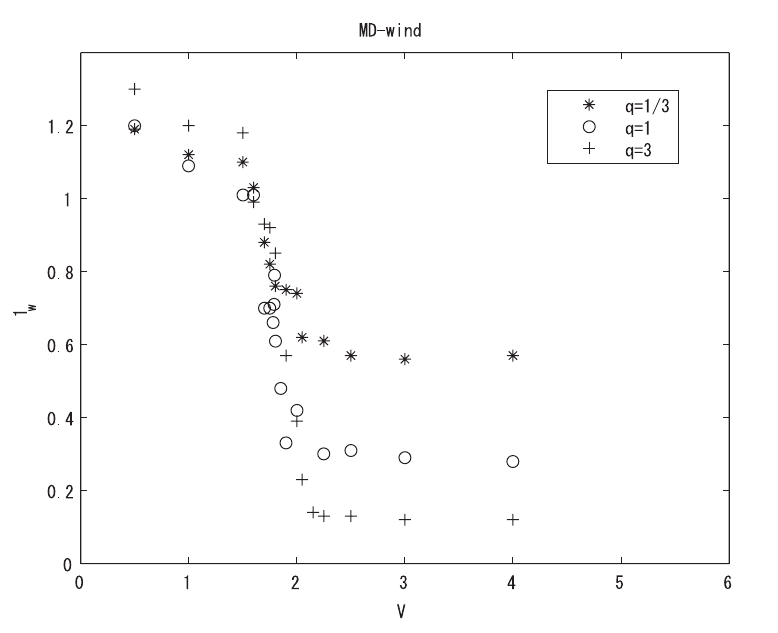}
   \caption{The specific angular momentum loss $l_w$ is  shown as a function of the outflowing velocity on the surface of the mass-losing star, $V$. Three mass ratios, $q = 1/3$, 1, and 3, are examined. The specific angular momentum loss  decreases from about 1--1.2 at low $V$ (RLOF) to 0.1--0.5 at high $V$ (wind type flow). For very fast wind, i.e., when $V > 2$, all the cases approach an asymptotic value. Reproduced with permission from Jahanara et al. \cite{Jaha05}.}
\label{boffig:lw}
\end{center}
   \end{figure}

The isotropic mass loss assumes that $l_w \equiv l_{w,s}=1/(1+q)^2$, easily derivable from assuming that all the mass lost carries the angular momentum of the mass-losing star. 
Jahanara et al. found (Fig.~\ref{boffig:lw}) that for high wind velocities one reaches such limiting value for mass ratios smaller than one, they obtain larger values than predicted -- by a factor two for $q=3$ and a factor of five for $q=7$. For 
low wind velocities, moreover, the specific angular momentum does not vary greatly as a function of the mass ratio and is much higher than the value predicted by this simple formula.
They obtain the following fit for the specific angular momentum loss : 
 \begin{eqnarray}
 l = 0.25 + \frac{0.12}{v_R+0.02} &  ~~~~~~~~~~q=1 \\
 l = 0.12 + \frac{0.12}{v_R+0.02} &  ~~~~~~~~~~q=3,
 \label{Eq:Jah}
 \end{eqnarray}
where $v_R$ is the velocity of the gas at the Roche radius.

The possibility that the specific angular momentum loss can be much larger than the canonical value has important consequences on the variations of the orbital parameters.  
 We can write the variation of the semi-major axis $\dot a$ as a function of the angular momentum loss $\dot J$ (for $e$=0): 
 \begin{equation}
 \frac{\dot a}{a} =  2 \frac{\dot J}{J} + \frac{\dot M_1+\dot M_2}{M_1+M_2}
- 2 \frac{\dot M_1}{M_1} - 2 \frac{\dot M_2}{M_2}.
\label{bofeq:jdot}
\end{equation}

Writing the angular momentum loss, $\dot J$, in function of the dimensionless specific angular momentum loss, $l_w$, i.e.  ${\dot J} = l_w
a^2 \Omega (\dot M_1+\dot M_2)$, we have: 

\begin{equation}
\frac{\dot a}{a} =  \frac{\dot M_1 + \dot M_2}{M_1+M_2} + 
2\frac{\dot M_1}{M_1} \left( l_w - 1 + l_w \frac{M_1}{M_2} \right) +
2\frac{\dot M_2}{M_2} \left( l_w - 1 + l_w \frac{M_2}{M_1} \right).
\end{equation}

Neglecting the mass accreted, we have 
\begin{equation}
\frac{\dot a}{a} = \left( \frac{M_1}{M_1 + M_2} + 2 \frac{M_1 + M_2}{M_2}
l_w - 2 \right) \frac{\dot M_1}{M_1} \equiv L(q)\frac{\dot{M}_{1}}{M_{1}}.
\label{bofeq:jaha6}
\end{equation}

With a value of  $l_{w}\sim 1$, the function in front of  $\dot{M}_{1}/M_{1}$ would be positive, hence the orbit would shrink, because  $\dot{M}_{1}<0$. As seen above,  $l_{w}\approx 1$ is a plausible value for low wind speed flows.
From Eq.~\ref{bofeq:jaha6}, 
 \begin{equation}
 L(q)=\frac{q}{1+q}+2 l_w (1+q)-2,
\end{equation}
 so that for the system to shrink we need to have $L(q)>0$ and, hence,  
$$l_{w}>\frac{1}{2} \frac{2+q}{(1+q)^{2}}.$$ 
From the results of the numerical simulations of \cite{Jaha05}, as shown by Eq.~\ref{Eq:Jah}, it appears that the system would shrink -- when assuming no mass accretion -- when the wind speed at the Roche lobe is smaller than about one ($q=1$) and three ($q=3$) times the orbital velocity. One can also show that the conclusion still holds, when the mass ratio is larger or equal to one, when mass accretion is taken into account. This makes the symbiotic channel for Type Ia supernovae a plausible one and could also dramatically help explain the existence of Barium stars and other Peculiar Red Giants with orbital periods below, say, 1000 days (see, e.g., \cite{Abate13}).

It still remains to be seen what would be the evolution of the eccentricity caused by this specific angular momentum loss, but in any case, it seems clear that binary population synthesis codes should take these corrected values for mass accretion and angular momentum loss into account when trying to explain post-AGB systems.

\section{The zoo of Peculiar Stars}

\subsection{Barium and related stars}

Barium stars\index{barium star} constitute about 1\% of G-K giants that show in their spectra a very strong Ba{II} 4554 line. The group was defined by \cite{BK51} from inspection of low resolution spectrograms. These authors also called attention to some other prominent features in the spectra of those stars, namely the enhancement of all BaII lines, of the 5165 C$_2$ band, as well as of the G-band due to CH\index{carbon band}. In addition, some bands of CN are also enhanced. An index of the strength of the BaII 4554 line has been introduced by \cite{Wa74}: the line strength is estimated visually on a scale from 1 (weak) to 5 (strong). Abundance analyses of the atmosphere of barium stars have shown that elements with Z $\leq$ 38 behave as in normal giants, except carbon which is enhanced. Heavier elements up to Ba (Z=56) are enhanced, and still heavier elements are generally not enhanced. 
This is the typical behaviour expected from the s-process\index{s-process}, one of the two processes of neutron captures on iron-peak elements. The s- (or slow-) process takes in particular place during the thermal pulses on the upper-Asymptotic Giant Branch (AGB\index{AGB star}). 

CH stars\index{CH star}\index{CH giant} present also strong bands of CH, the enhancement being larger than in barium stars. Most of the metallic lines are weakened, but heavy elements are enhanced. Other carbonated molecules, like C$_2$ and CN are also strengthened. 

In S stars\index{S Star}, the BaII lines are enhanced, as are the SrII lines\index{strontium}. However, S stars are of later type than the barium stars (around M0) and, therefore, exhibit ZrO bands\index{zirconium}. 

F str4077 stars were discovered by Bidelman \cite{Bid81} on the basis of objective-prism spectra. Those F-type main sequence stars have normal or deficient abundance of the iron-peak elements, but overabundance of the s-process elements. They are now commonly included, together with CH subgiants, in the class of dwarf barium stars \cite{Lu91}. Other carbon-rich dwarfs\index{carbon-rich dwarf} are also known (see, e.g., \cite{Gray01,JJ87,La85}). In barium stars, the overabundance of s-process elements over the solar value, ranges from a factor 2 (mild barium stars) to a factor 30 (strong barium stars). In CH stars, this enhancement is still larger, up to 100 sometimes. 

\begin{figure}
\begin{center} 
\includegraphics[width=119mm]{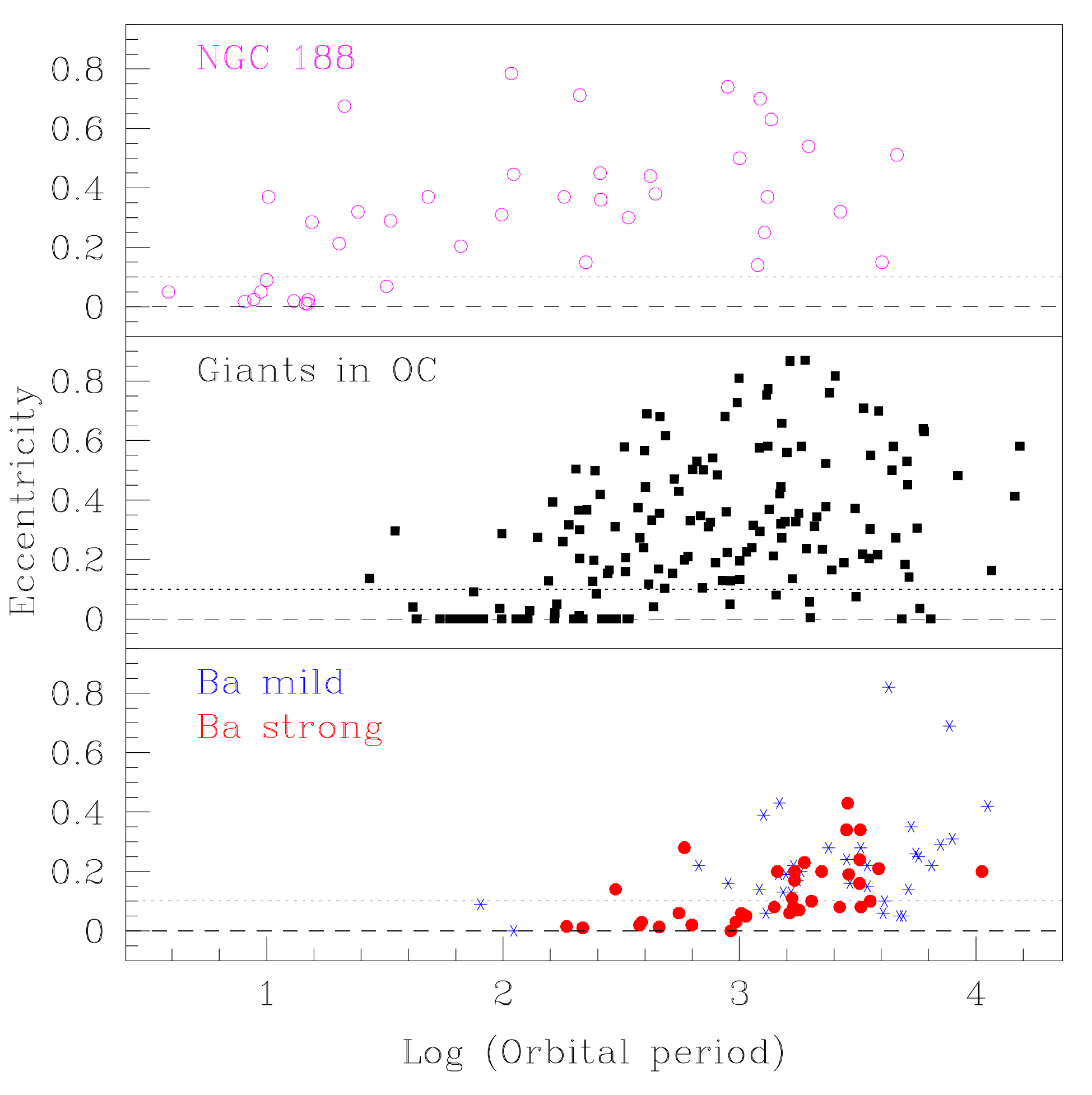}
\caption{Eccentricity-period (in days) diagrams for various samples. The upper panel shows the main-sequence stars in the open cluster NGC 188\index{NGC 188} (from \cite{Gel09}; see also Chap. 3); the middle panel shows red giant spectroscopic binaries in open clusters (from \cite{Merm}); while the lower panel shows a compilation of orbital elements for mild and strong barium stars. The dotted line shows the limit $e=0.1$.}\label{fig:BoffinelogP} 
\end{center}
\end{figure}

Thus barium and related stars present carbon and s-process element enhancements, which are typical of evolved AGB stars, but they are in an evolutionary stage prior to the AGB. This apparent paradox can be solved only by assuming that these peculiar stars are all binaries.
And indeed, in the early 80's, McClure, Fletcher \& Nemec \cite{McFN83} found that barium stars seemed to belong to binary systems, a conclusion further confirmed by \cite{JB92,JoM92,Jo98} and \cite{McW90}. These systems have orbital periods in the range 80 days to several years. Jorissen \& Mayor \cite{JoM92} also found that non-Mira S stars are predominantly binary systems with similar periods. The companion in these systems are all thought to be white dwarf, i.e., the remnant of a former AGB \cite{BoNP84,We86}. 
As clearly emphasised by Eggleton \cite{Egg92},
\begin{quotation}
Barium stars have a claim to be considered the most important class of binaries to have emerged in the last few decades. This is partly because 
their putative progenitors are intrinsically common, and partly because they themselves are quite common, i.e. they appear to have to be the fairly normal outcome of evolution in a binary with period in the range 1--10~yrs.
\end{quotation}

A striking property of barium stars is that they have non-circular orbits, even though they have eccentricities smaller than normal systems \cite{BPC92,BCP93}. 
Figure~\ref{fig:BoffinelogP} shows the eccentricity-orbital period diagrams for several samples of stars. The upper panel shows the main sequence stars in the old open cluster NGC 188 (Chap. 3) and is typical of main sequence spectroscopic binaries. It shows that such systems have almost any eccentricity at a given period with 2 exceptions: the closest of all systems have been circularised by tidal forces, while for periods above $\sim 50$~days, there are no binary systems with eccentricity below 0.1 (dotted line). When looking at the sample of red giants in open clusters \cite{Merm}, we can see that all the shortest period systems have disappeared due to RLOF -- they will have undergone mass transfer and most likely a common
envelope phase (see Chap. 8), and the red giant is now stripped of its mass and appears as a (He) white dwarf. The circularisation period has increased given the larger radii of red giants and, although there is a general trend for longer systems to be more eccentric, the eccentricity is quite large on average. A few systems occupy the large $P$ -- small $e$ region in the plane, and this is most likely because these are post-mass transfer systems, containing a red giant and a white dwarf. The third panel shows the diagram for barium stars and it is clear that although \emph{most systems are not circularised}, the average eccentricity is much smaller than for normal giants. Several systems also occupy the large $P$ -- small $e$ region. The fact that the orbits of barium stars are generally not circularised may indicate that these systems avoided a phase of RLOF, or that some eccentricity-pumping mechanism was excited after the RLOF. 
It is very interesting to compare the $e-\log P$ diagram of barium stars with that of blue straggler stars in NGC 188, as shown in Fig. 3.2 in Chap. 3. The striking resemblance may point to a common origin for these blue straggler stars and barium stars. In this respect, it would be useful to know if \emph{any of the blue straggler stars in NGC 188 are s-process and/or carbon enriched}.

Boffin \& Jorissen \cite{BJ88} devised a model of wind accretion from an AGB star to explain barium and related stars. They assumed a binary system with a wide enough orbit  in which the AGB star is transferring mass via a stellar wind to a main sequence companion which becomes polluted in carbon and s-process elements. The AGB then evolves into a white dwarf while the companion will appear as a carbon or barium dwarf  and, when on the giant branch, as a peculiar red giant. This model seems to reproduce the bulk properties of bariums stars \cite{BZ94,JB92}, although there is still a debate in the literature as to how these systems avoided RLOF and have their current non-zero eccentricity. The discussion above as shown, however, that it is possible to end up with the current orbital periods, even if the progenitors systems were much wider, and several studies have also shown that it is possible to reproduce the eccentricity distribution of barium stars \cite{Bona08,KTL00}.
 
The important features of the Boffin \& Jorissen wind accretion model is that it is able to explain:
\begin{itemize}
\item the existence of barium stars with very long orbital periods;
\item the existence of a weak correlation between s-process enhancement and orbital period;
\item the non-circular orbits of long period systems; and
\item the existence of dwarf barium stars.
\end{itemize}

Boffin \& Jorissen assumed the Bondi-Hoyle mass accretion rate, Eq.~\ref{Eq:BHbin} (in fact, they implemented the idea!). Boffin \& Zacs \cite{BZ94} showed, however, that even if the mass accretion rate was lower than that derived by this approximate formula, i.e., of a few percents only as shown above, it was still enough to explain the pollution of most barium stars. 
The initial model of Boffin \& Jorissen \cite{BJ88} fails, however, to explain most small period barium star systems with orbital period below, say, 1000 days. The existence of such systems was always a problem  since if one takes the value of the specific angular momentum given by the spherically symmetric wind, one finds that the system will widen. Things changes when one takes into account the specific angular momentum loss as computed by Jahanara et al. \cite{Jaha05}.
If the present system has a period below 1000 days, say, then for typical masses in a barium system, its present separation is roughly 2.5 AU, implying with the canonical value of $l_{w,s}$ an initial semi-major axis of less than 1 AU. This is too small to hope to fit an asymptotic giant branch star without having Roche-lobe overflow. With the value of the specific angular momentum loss found by  \cite{Jaha05}, the system did not expand, but instead shrank by a factor 3 to 10, depending on the ratio of the wind speed to the orbital velocity. Interestingly, Brookshaw \& Tavani \cite{BrookTav93} also emphasise the importance of angular momentum loss in the evolution of binary systems and the fact that the often used approximation $l_w=l_{w,s}$ can lead to wrong conclusions, in particular in X-ray binaries\index{X-ray binary}.

\subsection{Symbiotic stars and the case of SS Lep}

Symbiotic stars\index{symbiotic star} are a class of bright, variable red giant stars\index{red giant}, whose composite spectrum present typical absorption features of a cool star on top of strong hydrogen and helium emission lines\index{emission line}, linked to the presence of a hot star and a nebula. It is now well established that such a ``symbiosis'' is linked to the fact that these stars are active binary systems, with orbital periods between a hundred days and several years. Symbiotic stars are de facto the low- and intermediate-mass interacting binaries with the longest orbital periods and the largest component separations. They are thus excellent laboratories to study a large range of very poorly understood physical processes \cite{Mik07}. 

One of the main questions related to symbiotic stars is how the mass transfer takes place: by stellar wind\index{stellar wind}, through Roche-lobe overflow\index{Roche lobe overflow}  or through some intermediate process? Answering this question indeed requires being able to compare the radius of the stars to the Roche lobe radius. Optical interferometry can achieve this and in particular Blind et al. \cite{Blind11} used this unique ability to study in unprecedented detail the interacting binary system SS Leporis.

SS Leporis\index{SS Leporis} is composed of an evolved M giant and an A star in a 260-day orbit, and presents the most striking effect of mass transfer, the ``Algol''\index{Algol system} paradox; that is, the M giant is less massive than the A star. The latter is unusually luminous and surrounded by an expanding shell, certainly as the result of accretion. The observation of regular outbursts and of ultraviolet activity from the A star shell are further hints that mass transfer is currently ongoing. 
  
SS Lep was observed during the commissioning of the PIONIER\index{PIONIER} visitor instrument at the Very Large Telescope Interferometer\index{Very Large Telescope Interferometer}. Blind et al. were able to detect the two components of SS Lep as they moved across their orbit and to measure the diameter of the red giant in the system ($\sim2.2$ milli-arcseconds). By reconstructing the visual orbit and combining it with the previous spectroscopic one, they could further constrain the parameters of the two stars. 

The M giant is found to have a mass of 1.3 M$_\odot$, while the less evolved A star has a mass twice as large: thus a clear mass reversal must have taken place, with more than 0.7 M$_\odot$ having been transferred from one star to the other. The results also indicate that the M giant only fills around $85 \pm 3$\% of its Roche lobe, which means that the mass transfer in this system is most likely by a wind and not by RLOF. It is useful to note that given the rather low mass loss from the giant in SS Lep, the Roche lobe will not be much affected, as the $f$ parameter (cf. Sec. \ref{Boffinsec:1}) is most likely only of order 0.01--0.1, leading to a decrease of the Roche lobe by 1 to 4\% only. 

Podsiadlowski \& Mohamed \cite{PM07} suggest the possibility of a new mode of mass transfer -- the wind Roche lobe overflow\index{wind Roche lobe overflow} -- where a slow wind fills the Roche lobe. This is in effect the low wind velocity case of Nagae et al. \cite{Nagae04}; see Fig.~\ref{nagae}. Because the wind speed in M giants is rather small (around 10--15 km/s) and lower than the orbital one for SS Lep ($v_{\rm orb}$ = 48 km/s), we expect it to be in the particular case of a wind Roche lobe overflow, where a substantial part of the stellar wind can be accreted. The simulations of Nagae et al. show that at least 10\% of the M giant wind could be accreted in SS Lep. Abate et al. \cite{Abate13} find much larger mass transfer, up to 50\%. The mass transfer cannot be fully conservative though, as matter will generally also escape through the outer Lagrangian point\index{Lagrangian point}.

However, it is still difficult to explain the current system state with normal stellar-wind rates, which are too low. Indeed, before the AGB phase, the typical mass-loss rates are around $\sim 1-2 \times 10^{-8}$~M$_\odot$yr$^{-1}$ at the normal (e.g., Reimers) rate\index{stellar wind}. The M giant (with an expected initial mass $> 2.2$~M$_\odot$) should have lost only a few hundredths of a solar mass before reaching the AGB, whereas we expect it to have lost at least 0.9~M$_\odot$. As the M star is on the AGB since only a few million years -- and will stay there for a few million years more at most -- it cannot have lost much mass since then.
There is, however, some evidence of enhanced wind mass loss of giants in binaries compared to single giants of the same spectral type \cite{Jorissen03,Mik07}. From a theoretical point of view, the presence of a companion reduces the effective gravity of the mass-losing star, thus enhancing the mass loss. In the case of SS Lep, the superbly called companion-reinforced attrition process\index{companion-reinforced attrition process} (``CRAP'') of tidally enhanced stellar wind \cite{TE88} allows a mass loss rate 150 times higher%
\footnote{The CRAP supposes that the typical mass loss, $\dot M_R$, is tidally enhanced by the companion and Tout \& Eggleton \cite{TE88} propose the following \emph{ad hoc} formula:
\begin{equation}
 \dot M = \dot M_R ~\left \{ 1 + B \times min \left[ \left(\frac{R}{R_L}\right)^6,\frac{1}{2^6} \right] \right \},
 \end{equation}
with $B \approx 10^4$. It is important to note that the existence of many different kinds of binary systems seem to require the CRAP to work, with different values of $B$, however.} 
than the Reimers rate, i.e. $\sim 2.4 \times 10^{-6}$~M$_\odot$yr$^{-1}$. 
Quite noteworthy, Bona{\v c}i{\'c} Marinovi{\'c}, Glebbeek, \& Pols \cite{Bona08} and Karakas, Tout \& Lattanzio \cite{KTL00} also had  to invoke some tidally-enhanced mass loss\index{tidally-enhanced mass loss} from the AGB star to provide an explanation for
the eccentricities of most barium star\index{barium star} systems. And so did Abate et al. \cite{Abate13} to explain carbon-enhanced metal-poor stars.

Blind et al. were able to reproduce the current state of SS Lep, when including wind RLOF and the CRAP. They started with a system having an initial period of 160 days and initial masses $M_{M,0}$ = 2.28~M$_\odot$, $M_{A,0}$ = 1.85~M$_\odot$. For about 1 Gyr, the system evolves without much change, and the primary star starts its ascent of the AGB. After 2.8 Myr, the masses and period have reached the currently observed values, with about 0.1~M$_\odot$ having been lost by the system, and forming some circumbinary disc. The mass loss and transfer occurred mostly during the last 500~000 years, with a mass loss $\sim 10^{-6}$~M$_\odot$yr$^{-1}$. During the whole process, the Roche lobe radius around the M star remained similar, the lowest value being 74~R$_\odot$. No RLOF should thus have happened, unless the initial eccentricity was very high. More detailed simulations (L. Siess, priv. comm.) seem to question the exact details of this scenario, but it should be working in a broad sense.

Boffin et al. \cite{Bo14} conducted a mini-survey of symbiotic stars with PIONIER and found that for the three stars with the shortest orbital periods (i.e., HD 352\index{HD 352}, FG Ser\index{FG Ser} and HD 190658\index{HD 190658}), the giants are filling (or are close to filling) their Roche lobes, while the other three studied stars (V1261 Ori\index{V1261 Ori}, ER Del\index{ER Del}, and AG Peg\index{AG Peg}) have filling factors in the range 0.2 to 0.55, i.e., the star is well within its Roche lobe. They also tentatively propose \cite{BoMes14} that the systems which apparently fill their Roche lobes\index{Roche lobe} are those that contain a main-sequence companion or a helium white dwarf (WD)\index{white dwarf}, and not a carbon/oxygen (CO) WD.

      \begin{figure}
\begin{center} 
   \includegraphics[width=119mm]{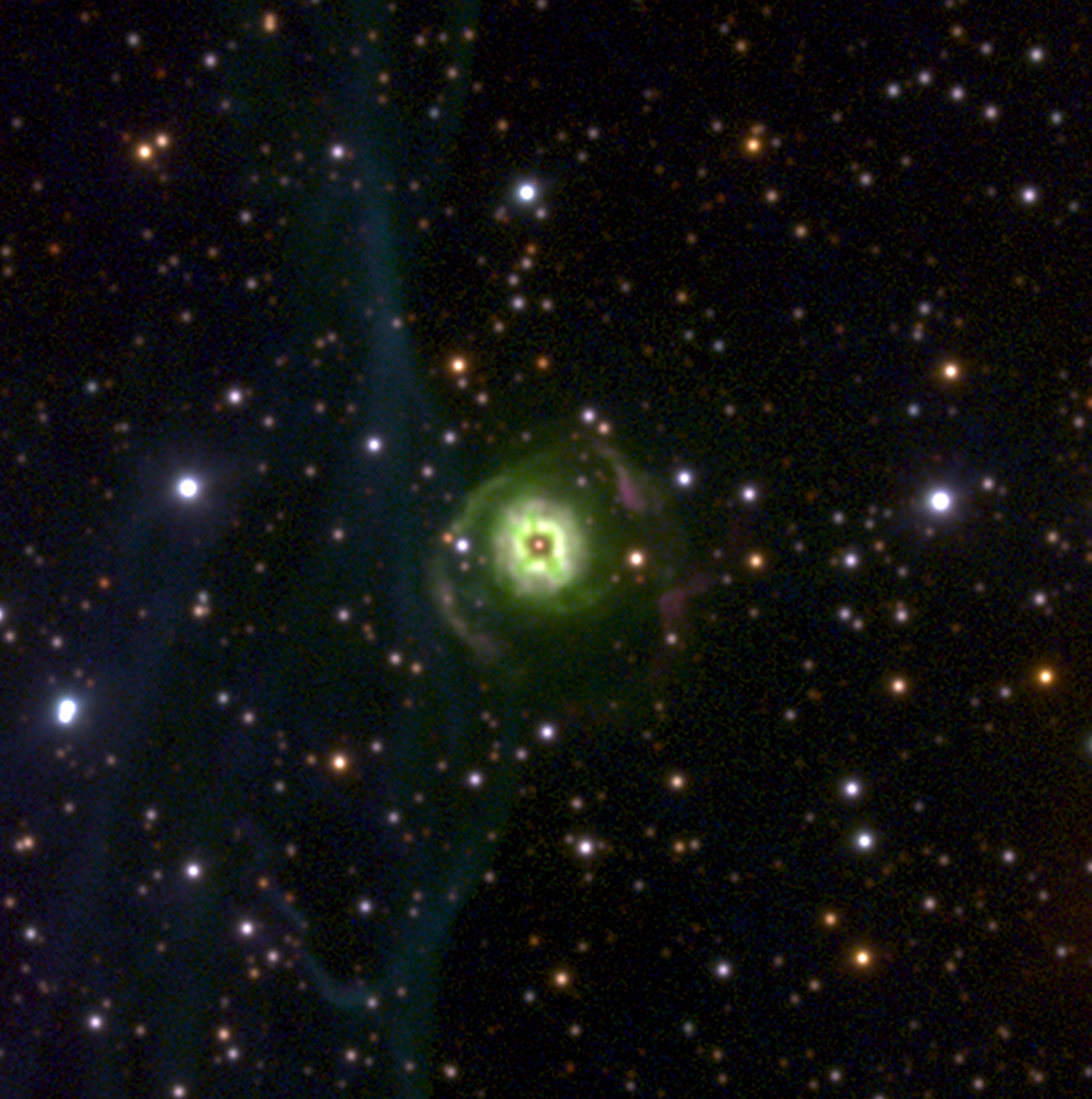}
   \caption{ VLT FORS2 colour-composite image of Hen 2-39 made from H$\alpha$+[N II] (red), [O III] (green) and [O II] (blue). Reproduced with permission from 
\cite{Hen239}.}
\label{boffig:hen2-39}
 \end{center}
  \end{figure}

\subsection{The Fellowship of the Ring}
\label{bofsec:ring}
A small group of planetary nebulae\index{planetary nebula} (PNe) are known to harbour binary central stars where a subgiant or giant companion is enriched in carbon and s-process elements (see \cite{BoMi12,2012MNRAS.419...39M} and ref. therein). These are the progenitors of the barium stars, caught during the short-lived phase ($\sim 10^4$~yr) when the PN ejected by the white dwarf is still visible. 
These 
cool central stars are those of the (unfortunately) denominated  A~35-type \cite{1993IAUS..155..397B}. The initial list included A~35\index{Abell 35}, LoTr~1\index{LoTr~1} and LoTr~5 which have rapidly rotating sub-giants or giants accompanied by very hot white dwarfs peaking at UV wavelengths ($T_\mathrm{eff}\geq100$ kK). Since these initial discoveries, Bond, Pollacco \& Webbink \cite{2003AJ....125..260B}  added WeBo~1 to the list while Frew \cite{2008PhDT.......109F} raised suspicions that the nebula of A~35 may not be anymore a bona-fide PN, even if it is likely to have passed through a PN phase at some stage to create the stellar abundances currently observed (hence, the group should be named differently!). Further additions of HD~330036\index{HD~330036} and AS~201\index{AS~201} may be drawn from barium enhanced yellow or D'-type symbiotic stars\index{symbiotic star} which show extended nebulae, but the classification of such objects as PNe is controversial \cite{c03,2005A&A...441.1135J}. On the other hand, Tyndall et al. \cite{Ty13} showed that the cool central star of LoTr~1 is not enhanced in barium and is therefore not a real member of the group. It is also interesting to note that van Winckel \cite{vW14} reported  the first detection of orbital motion in LoTr5\index{LoTr5}. Despite a 1807-day coverage, a full period was not yet covered, indicating that the orbital period may be as much as 10 years. This makes for a direct link with barium stars.

Miszalski et al. \cite{2012MNRAS.419...39M} present UV and optical observations of the \emph{diamond ring} PN A~70\index{PN A~70} (PN G038.1$-$25.4\index{PN G038.1$-$25.4}) that prove the existence of a barium star binary CSPN. The apparent morphology of A 70 is that of a ring nebula, although on closer inspection the [O~III] image shows a ridged appearance similar to Sp 1\index{Sp 1} which is a bipolar torus viewed close to pole-on. 

More recently, Miszalski et al. \cite{Hen239} also showed that the cool central star of the planetary nebula Hen 2-39\index{Hen 2-39} (Fig.~\ref{boffig:hen2-39}) is carbon and s-process enriched. 
Weak Ca~II K emission is detected, indicating some chromospheric activity may be present. Time-resolved photometry revealed a short-term periodicity,  indicating that the  nucleus of Hen 2-39 is a fast rotating giant (with a period of a few days), consistent with the periods found for similar stars. These include 5.9~d for LoTr 5  and 4.7~d for WeBo 1. A rotation period of 6.4~d was also determined for LoTr 1 \cite{Ty13}. Given our discussion in Sect.~\ref{bofsec:rot}, one could conjecture that these stars were spun-up during a wind accretion phase.

A significant carbon enhancement of [C/H] = 0.42 $\pm$ 0.02?dex is required to fit the observed spectrum of Hen 2-39. The iron abundance was found to be close to solar while barium is clearly overabundant, [Ba/Fe] = 1.50 $\pm$ 0.25~dex. Miszalski et al. \cite{Hen239} found that it was possible to reproduce such pollution, using 1.8~M$_\odot$ as the initial mass for the AGB star and assuming that the polluted star accreted about 0.5~M$_\odot$ from it during the last thermal pulses. This implies quasi-conservative mass transfer at this stage.

The nebula shows an apparent ring-like morphology as also seen in the other PNe with barium CSPNe, WeBo 1 and A 70. 
It seems therefore clear that the morphology must be the outcome of the mass transfer episode -- most likely by wind -- that led to the pollution of the cool central star in carbon and s-process elements. As such, LoTr~5, WeBo~1, A~70, and Hen~2-39 belong to a \emph{Fellowship of the Ring}, all progenitors of barium stars. Their further study should bring detailed knowledge on the wind mass transfer. Moreover, the short time spent during the PN phase makes barium CSPNe a potentially very powerful platform for studying AGB nucleosynthesis in that we simultaneously see both the polluted s-process-rich cool star and the nebula ejected by the polluting star. 

\subsection{Evidence for wind accretion before common-envelope evolution}
\index{common-envelope evolution}

The existence of barium central stars to planetary nebulae\index{planetary nebula} constitutes the only firm evidence for mass transfer and accretion onto non-WD companions in PNe. Observations of close binary central stars, i.e. those that have passed through a common-envelope (CE) phase and have orbital periods less than $\sim1-2$~d (e.g., \cite{DM08,Mis09}), show no evidence for rapid variability (flickering) or spectroscopic features that could be attributed to accretion.

The presence of collimated outflows or jets surrounding several systems (e.g., \cite{Bo12})  possibly launched from  an accretion disc\index{accretion disc} that is no longer present may be indirect evidence for accretion, either prior to the common envelope phase via wind accretion from the AGB primary, during the start of the CE infall phase or perhaps even after the CE phase.  
Long-slit observations of some jets\index{jet} around post-CE nebulae indicate the jets were probably ejected before their main nebulae \cite{Bo12,Cor11,Mitchell07,Mis11}. Another fundamental clue comes from point-symmetric outflows of PNe. Simulations can recreate these complex outflows with a precessing accretion disc around the secondary launching jets (e.g., \cite{Cliffe95,Raga09}). While there are multiple examples of such PNe, none were known to have a binary nucleus until the landmark discovery of a post-CE binary nucleus in the archetype of this class Fleming 1\index{Fleming 1} \cite{Bo12}. The characteristics of Fleming 1 indicate that wind accretion (with the formation of a disc) must have happened \emph{before} a CE episode.

However, the clearest proof for accretion would be a polluted main-sequence companion with an atmosphere strongly enriched by accreted material. 
Miszalski, Boffin \& Corradi \cite{2013MNRAS.428L..39M}  report  the detection of a carbon dwarf secondary in the post-CE central star binary in the Necklace\index{The Necklace} planetary nebula (PN G054.2$-$03.4\index{PN G054.2$-$03.4}) as the first firm proof for a previous accretion phase. These authors found that to reproduce the observed carbon enhancement, they need to accrete between 0.03 and 0.35~M$_\odot$, depending on the mass of the star, in a binary system with initial orbital period between 500 and 2000 days. The current period of the system, 1.16 d, clearly proves that the system underwent a common-envelope phase. 
The most advanced simulations of the spiral-in part of the CE phase by Ricker \& Taam \cite{RT12} and Passy et al. \cite{Passy12} predict a negligible amount of mass accretion $10^{-3}$~M$_\odot$. The accretion is therefore most likely to have occurred before the CE phase via wind accretion, a process that simulations predict to form an accretion disc around the companion (e.g. \cite{TBJ96}). Since the jets of the Necklace are also observed to be older than the main PN \cite{Cor11}, they were probably launched from such a disc. 

The initial period estimated for the system is typical of symbiotic stars, where substantial wind accretion on to companions is known to occur. A small CE efficiency would then produce a dramatic decrease in the orbital period to the current value.
The Necklace nebula is therefore a real {\bf Rosetta Stone} for the study of mass transfer.

\begin{acknowledgement}
It is a pleasure to thank my many collaborators over the years for exciting work on this topic: A. Acker, 
U. Anzer, 
N. Blind, 
J.-P. Berger,
N. Cerf, R.~L.~M. Corradi, H. Fujiwara, 
I. Hachisu, H. Isaka, 
B. Jahanara, D. Jones, A. Jorissen,
J.-B. Le Bouquin, 
 T. Matsuda, B. Miszalski,
M. Mitsumoto, T. Nagae, K. Oka, 
 G. Paulus,
 E. Shima,
T. Theuns,
 A.~A. Tyndall, and
L. Za{\v c}s
\end{acknowledgement}

\backmatter
\printindex


\end{document}